\documentclass[aps,prd,preprint,nofootinbib,11pt]{revtex4}
\usepackage{amsfonts}
\usepackage{mathrsfs}
\usepackage{graphicx}
\usepackage{amsmath}
\usepackage{amssymb}
\usepackage{multirow}
\usepackage{subfigure}
\usepackage{epsfig}
\usepackage{graphicx}
\usepackage{color}
\newcommand{\bqa}{\begin{eqnarray}}
\newcommand{\eqa}{\end{eqnarray}}
\newcommand{\beq}{\begin{equation}}
\newcommand{\eeq}{\end{equation}}
\allowdisplaybreaks[1]
\graphicspath{{fig/}{dia/}} \DeclareGraphicsExtensions{.eps}

\begin{document}

\title{Fully Charmed Tetraquark States in $8_{[c\bar{c}]}\otimes8_{[c\bar{c}]}$ Color Structure via QCD Sum Rules\\[0.7cm]}
\author{Chun-Meng Tang$^{1}$,  Chun-Gui Duan$^{1,2}$\footnote{duancg@hebtu.edu.cn}, Liang Tang$^{1}$\footnote{tangl@hebtu.edu.cn}}

\affiliation{$^1$ College of Physics and Hebei Key Laboratory of Photophysics Research and Application,
Hebei Normal University, Shijiazhuang 050024, China \\
$^2$ Hebei Key Laboratory of Physics and Energy Technology, North China Electric Power University, Baoding 071000, China
}

%\author{~\\~\\}

%\affiliation{}

\begin{abstract}
\vspace{0.3cm}
Stimulated by the recent experimental results on the fully-charm tetraquark states, we systematically calculate the mass spectra of the fully-charm tetraquark states in $8_{[c\bar{c}]}\otimes8_{[c\bar{c}]}$ color configuration via QCD sum rules. By constructing nine $8_{[c\bar{c}]}\otimes8_{[c\bar{c}]}$ type currents with quantum numbers $J^{PC}=0^{-+},0^{--},1^{-+},1^{+-},1^{--}$ and $2^{++}$, we perform analytic calculation up to dimension six in the Operator Product Expansion (OPE). We find the fully-charm tetraquark states with $J^{PC}=1^{+-},2^{++}$ lie around 6.48 $\sim$ 6.62 GeV while the fully-charm tetraquark states with $J^{PC}=0^{-+},0^{--},1^{--},1^{-+}$ are about 6.85 $\sim$ 7.02 GeV. Notably, the mass predictions for the $c\bar{c}c\bar{c}$ tetraquarks, specifically those with $J^{PC}=2^{++}$, align with the broad structure identified by LHCb. Moreover, the masses of fully-charm tetraquarks with $J^{PC}=0^{-+}$ and $1^{-+}$ are anticipated to match closely with the mass of X(6900), considering the margin of error. Such findings hint at the presence of some $8_{[c\bar{c}]}\otimes8_{[c\bar{c}]}$ components within the di-$J/\psi$ structures observed by LHCb. The predictions for tetraquark states with $J^{PC}=0^{--},1^{+-},1^{--}$ may be accessible in the future BelleII, Super-B, PANDA, and LHCb experiments.
\end{abstract}
\maketitle
\newpage

\section{Introduction}
In 1964, Gell-Mann \cite{Gell-Mann:1964ewy} and Zweig \cite{Zweig:1964ruk} proposed the concept of exotic hadrons possessing quark compositions beyond the minimal configurations ($q\bar{q}$ or $qqq$), such as tetraquark states, hybrid states, and glueball, etc. These configurations are also consistent with the principles of Quantum Chromodynamics (QCD). Following the discovery of X(3872) in 2023~\cite{Belle:2003nnu}, the wealth of experimental data accumulating over time has led to the identification of numerous multiquark states through research conducted at BelleII, BESIII, LHCb, etc. It goes without saying that delving into the characteristics of multiquark exotic states represents a particularly fascinating subject within hadronic physics, increasingly drawing the attention of both theorists and experimentalists.

The LHCb Collaboration has reported resonant structures in the double J/$\psi$ invariant mass spectrum in 2020. This analysis utilized proton-proton collision data collected at center-of-mass energies of 7, 8 and 13 Tev from the LHCb experiments, corresponding to an integrated luminosity of 9$fb^{-1}$~\cite{LHCb:2020bwg}. They observed a broad structure beyond the threshold ranging from 6.2 to 6.8 GeV, as well as a narrow structure at approximately 6.9 GeV, named X(6900). Recently, the ATLAS and CMS Collaborations have verified the presence of X(6900) within the $J/\psi$$J/\psi$ invariant mass spectrum, achieving a statistical significance well beyond 5$\sigma$~\cite{CMS:2023owd,ATLAS:2023bft}. Moreover, in the same channel, ATLAS identified two additional structures, X(6400) and X(6600), while CMS detected X(6600) and X(7200) under two distinct analyses, one accounting for and the other disregarding interference effects among the resonances.

The exploration of $cc\bar{c}\bar{c}$ states in theory dates back significantly~\cite{Iwasaki:1976cn}, with their systematic study initiated by Chao in 1981 using the quark-gluon model~\cite{Chao:1980dv}. This pioneering work marked the first time exotic hadrons containing $cc\bar{c}\bar{c}$ were analyzed, predicting their masses to fall between 6.4-6.8 GeV, all exceeding the threshold for two charmonia. Subsequent research, as referenced in studies~\cite{Ader:1981db, Badalian:1985es, Heller:1985cb}, delved further into fully-charm tetraquark states, suggesting these primarily decay into pairs of charmonia. Fully heavy tetraquark states have attracted significant attention in recent years and have been extensively studied in various model frameworks~\cite{Anwar:2017toa, Bedolla:2019zwg, Lloyd:2003yc, Barnea:2006sd, Debastiani:2017msn, Wu:2016vtq, Wang:2019rdo, Liu:2019zuc, Faustov:2020qfm, Lu:2020cns, Heupel:2012ua, Weng:2020jao, Berezhnoy:2011xn, Karliner:2016zzc, Berezhnoy:2011xy, Feng:2020riv, Zhang:2020hoh, Karliner:2020dta, Wang:2020wrp, Giron:2020wpx, Chao:2020dml, Maiani:2020pur, Richard:2020hdw, Zhu:2020xni, Guo:2020pvt, Maciula:2020wri, Zhu:2020snb, Eichmann:2020oqt, Gong:2020bmg, Becchi:2020uvq, Dong:2020nwy, Chen:2016jxd, Wang:2017jtz, Chen:2018cqz, Wang:2018poa, Zhang:2020xtb, Wang:2020dlo, Albuquerque:2020hio, Chen:2020xwe, Wang:2021mma, Wan:2020fsk, Agaev:2023wua, Agaev:2023gaq, Agaev:2023ruu, Agaev:2023rpj, Agaev:2023ara, Jin:2020jfc, Zhao:2020nwy, Wang:2020ols, Wang:2021kfv, Liu:2021rtn, Mutuk:2021hmi, Dong:2021lkh, Wang:2022xja, Wang:2022jmb, Gong:2022hgd, Yu:2022lak, Wang:2021wjd, Liu:2020tqy}. Among the various approaches, the QCD sum rules method stands out for its unique benefits in investigating the properties of hadrons that involve non-perturbative QCD aspects. This technique has been applied in recent studies to examine the X(6900) particle, as cited in references~\cite{Chen:2016jxd, Wang:2017jtz, Chen:2018cqz, Wang:2018poa, Zhang:2020xtb, Wang:2020dlo, Albuquerque:2020hio, Chen:2020xwe, Wang:2021mma, Wan:2020fsk, Agaev:2023wua, Agaev:2023gaq, Agaev:2023ruu, Agaev:2023rpj, Agaev:2023ara}.

Given that the broad structure and X(6900) are both significantly above the $\eta_{c}\eta_{c}$ and $J/\psi J/\psi$ thresholds respectively, and lack light flavor quarks, it is improbable for these four-charm constructs to constitute hadronic molecules. This is because such molecules are created by exchanging light mesons, resulting in very low binding energies.  Consequently, the concept of a color-bound diquark-antidiquark configuration is commonly employed to elucidate the characteristics of the recently identified X(6900). It's important to highlight that Quantum Chromodynamics (QCD) suggests an alternative tetraquark structure $[8]_{Q\bar{Q}}\otimes [8]_{Q\bar{Q}}$, consisting of two color-octet components, as indicated in studies~\cite{Latorre:1985uy, Narison:1986vw, Wang:2006ri, Wang:2015nwa, Tang:2016pcf, Tang:2019nwv}.
The presence of QCD interactions sets this structure apart from a molecular state composed of two color-singlet mesons, allowing it to potentially decay into two charmonia through the exchange of one or more gluons. This underscores the significance of investigating the color octet-octet tetraquark state in the discovery of potentially new exotic hadrons. In our previous work\cite{Yang:2020wkh}, we have studied the fully-heavy tetraquark state with the scalar ($J^{PC}=0^{++}$) $[8]_{Q\bar{Q}}\otimes [8]_{Q\bar{Q}}$ type currents. In this paper, we systematically calculate the mass spectra for full-charm $c\bar{c}c\bar{c}$ tetraquark states with various quantum numbers in terms of QCD sum rules.

The remainder of this paper is organized as follows: Following the introduction, Section II presents the primary formulas. Section III details the numerical analyses and findings. The paper concludes with a discussion and summary of the results in the final section.

\section{Formalism}
The foundation of QCD sum rules begins with the construction of a two-point correlation function, which is established between two hadronic currents~\cite{Shifman:1978bx, Shifman:1978by, Reinders:1984sr, Narison:1989aq, Colangelo:2000dp}. For scalar and pseudo-scalar currents, the correlation function is delineated as follows:
\begin{eqnarray}
\Pi(q)&=&i\int d^{4}xe^{iq\cdot x}\langle 0|T\{j(x),j^{\dagger}(0)\}|0 \rangle,
\end{eqnarray}
while for the vector and axial-vector currents:
\begin{eqnarray}
\Pi_{\mu\nu}(q)&=&i\int d^{4}xe^{iq\cdot x}\langle 0|T\{j_{\mu}(x),j_{\nu}^{\dagger}(0)\}|0 \rangle.
\end{eqnarray}
The correlation function $\Pi_{\mu\nu}(q)$ is composed of two parts: $\Pi_{V}(q^{2})$ symbolizes the invariant function for spin-1, while $\Pi_{S}(q^{2})$ denotes the invariant function for spin-0, which can be expressed as:
\begin{eqnarray}
\Pi_{\mu\nu}(q) = \left(-g_{\mu\nu} + \frac{q_{\mu}q_{\nu}}{q^{2}}\right)\Pi_{V}(q^{2}) + \frac{q_{\mu}q_{\nu}}{q^{2}}\Pi_{S}(q^{2}).
\end{eqnarray}
For the tensor currents, we have
\begin{eqnarray}
\Pi_{\mu\nu,\rho\sigma}(q) = i\int d^{4}xe^{iq\cdot x}\langle 0|T\{j_{\mu\nu}(x),j_{\rho\sigma}^{\dagger}(0)\}|0 \rangle.
\end{eqnarray}
The Lorentz structure of the two-point correlation function for the tensor current is characterized in a specific structure:
\begin{eqnarray}
\Pi_{\mu\nu,\rho\sigma}(q) = T_{\mu\nu,\rho\sigma}\Pi_{2}(q^{2})+\cdots,
\end{eqnarray}
where the function $\Pi_2(q^2)$ is associated with the intermediate state of spin-2, and the ``$\cdots$" represents other structures which are independent
of the correlation function $\Pi_2(q^2)$. The projection factor $T_{\mu\nu,\rho\sigma}$ is a unique fourth-order Lorentz tensor, formulated by $g_{\mu\nu}$ and $q_{\mu}$,
\begin{eqnarray}
T_{\mu\nu,\rho\sigma} = \frac{1}{2}\left[ g_{\mu\rho}^{t}(q)g_{\nu\sigma}^{t}(q) + g_{\mu\sigma}^{t}(q)g_{\nu\rho}^{t}(q) - \frac{2}{3}g_{\mu\nu}^{t}(q)g_{\rho\sigma}^{t}(q)\right],
\end{eqnarray}
where
\begin{eqnarray}
g_{\mu\nu}^{t}(q) = (g_{\mu\nu} - q_{\mu}q_{\nu}/q^{2}).
\end{eqnarray}
Interpolating currents for fully-charm tetraquark states, each possessing distinct quantum numbers $J^{PC}$, are constructed as follows:
\begin{eqnarray}
j^{0^{-+}}_{A}(x) &=& \left[ \bar{Q}^{j}(x)\gamma_{5}(t^{a})_{jk}Q^{k}(x) \right] \left[ \bar{Q}^{m}(x)(t^{a})_{mn}Q^{n}(x) \right]\; \label{current-1} ,\\
j^{0^{-+}}_{B}(x) &=& \left[ \bar{Q}^{j}(x)\sigma_{\mu\nu}(t^{a})_{jk}Q^{k}(x) \right] \left[ \bar{Q}^{m}(x)\sigma_{\mu\nu}\gamma_{5}(t^{a})_{mn}Q^{n}(x) \right]\; \label{current-2} ,\\
j^{0^{--}}(x) &=& \left[ \bar{Q}^{j}(x)\gamma_{\mu}(t^{a})_{jk}Q^{k}(x) \right] \left[ \bar{Q}^{m}(x)\gamma_{\mu}\gamma_{5}(t^{a})_{mn}Q^{n}(x) \right]\; \label{current-3} ,
\end{eqnarray}
\begin{eqnarray}
j^{1^{--}}_{A\mu}(x) &=& \left[ \bar{Q}^{j}(x)\gamma_{\mu}(t^{a})_{jk}Q^{k}(x) \right] \left[ \bar{Q}^{m}(x)(t^{a})_{mn}Q^{n}(x) \right]\; \label{current-4} ,\\
j^{1^{--}}_{B\mu}(x) &=& \left[ \bar{Q}^{j}(x)\gamma_{\alpha}\gamma_{5}(t^{a})_{jk}Q^{k}(x) \right] \left[ \bar{Q}^{m}(x)\sigma_{\alpha\mu}\gamma_{5}(t^{a})_{mn}Q^{n}(x) \right]\; \label{current-5} ,\\
j^{1^{+-}}_{\mu}(x) &=& \left[ \bar{Q}^{j}(x)\gamma_{\mu}(t^{a})_{jk}Q^{k}(x)\right]\left[ \bar{Q}^{m}(x) \gamma_{5} (t^{a})_{mn}Q^{n}(x) \right]\; \label{current-6} , \\
j^{1^{-+}}_{A\mu}(x) &=& \left[ \bar{Q}^{j}(x)\gamma_{5}(t^{a})_{jk}Q^{k}(x) \right] \left[ \bar{Q}^{m}(x)\gamma_{\mu}\gamma_{5}(t^{a})_{mn}Q^{n}(x) \right]\; \label{current-7} ,\\
j^{1^{-+}}_{B\mu}(x) &=& \left[ \bar{Q}^{j}(x)\gamma_{\mu}(t^{a})_{jk}Q^{k}(x) \right] \left[ \bar{Q}^{m}(x)\sigma_{\mu\nu}(t^{a})_{mn}Q^{n}(x) \right]\; \label{current-8} ,\\
j^{2^{++}}_{\mu\nu}(x) &=& \left[\bar{Q}^{j}(x)\gamma_{\mu}(t^{a})_{jk}Q^{k}(x)\right]\left[ \bar{Q}^{m}(x)\gamma_{\nu}(t^{a})_{mn}Q^{n}(x) \right]\; \label{current-9} ,
\end{eqnarray}
where the $j,k,m,n=1,2,3$ and $a=1,2, \cdots, 8$ are color indices, the $t^{a} = \lambda^{a}/2$ is the Gell-Mann matrix, Q represents the charm quark. The interpolating currents under discussion are formed from two color-octet $Q\bar{Q}$ components.

At the quark-gluon level, we compute the correlation function using the operator product expansion (OPE) technique. Our analysis employs the heavy-quark propagator $S_{jk}^{Q}(p)$ in momentum space, which is expressed as:
\begin{eqnarray}
S^Q_{j k}(p) \! \! & = & \! \! \frac{i \delta_{j k}(p\!\!\!\slash + m_Q)}{p^2 - m_Q^2} - \frac{i}{4}g_{s} \frac{t^a_{j k} G^a_{\alpha\beta} }{(p^2 - m_Q^2)^2} [\sigma^{\alpha \beta}
(p\!\!\!\slash + m_Q)
+ (p\!\!\!\slash + m_Q) \sigma^{\alpha \beta}] \nonumber \\ &-&
\frac{i}{4}g_{s}^{2}(t^{a}t^{b})_{jk} G^{a}_{\alpha\beta}G^{b}_{\mu\nu}\frac{(p\!\!\!\slash + m_Q)}{(p^2 - m_Q^2)^5}(f^{\alpha\beta\mu\nu}+f^{\alpha\mu\beta\nu}+f^{\alpha\mu\nu\beta}) (p\!\!\!\slash + m_Q)\nonumber \\ &+& \frac{i \delta_{j k}}{48} \bigg\{ \frac{(p\!\!\!\slash +
m_Q) [p\!\!\!\slash (p^2 - 3 m_Q^2) + 2 m_Q (2 p^2 - m_Q^2)] }{(p^2 - m_Q^2)^6}
\times (p\!\!\!\slash + m_Q)\bigg\} \langle g_s^3 G^3 \rangle \; ,
\end{eqnarray}
where the subscripts $j$ and $k$ denote the color indices of heavy quarks, with the vacuum condensates are clearly displayed. Here, we difine $f^{\alpha\beta\mu\nu}\equiv \gamma^{\alpha}(p\!\!\!\slash + m_Q)\gamma^{\beta}(p\!\!\!\slash + m_Q)\gamma^{\mu}(p\!\!\!\slash + m_Q)\gamma^{\nu}$. For additional details on the discussed propagators, readers are encouraged to consult Refs.~\cite{Reinders:1984sr, Wang:2013vex, Albuquerque:2012jbz}.

The correlation function $\Pi(q^{2})$ from the perspective of quark-gluon interactions can be expressed through the application of the dispersion relation
\begin{eqnarray}\label{Pi-OPE}
  \Pi^{\text{OPE}}(q^2) = \int_{(4 m_Q)^2}^\infty ds \frac{\rho^{\text{OPE}}(s)}{s - q^2} + \Pi^{\langle GG \rangle}(q^2) + \Pi^{\langle GGG \rangle}(q^2),
\end{eqnarray}
where $\rho^{\text{OPE}}(s) = \text{Im} [\Pi^{\text{OPE}}(s)]/\pi$ and
\begin{eqnarray}\label{rho-OPE}
  \rho^{\text{OPE}}(s) &=& \rho^{\text{pert}}(s) + \rho^{\langle GG \rangle}(s)+ \rho^{\langle GGG \rangle}(s).
\end{eqnarray}
Within Eq.~\eqref{Pi-OPE}, the second term $\Pi^{\langle GG \rangle}(q^2)$ and the third term $\Pi^{\langle GGG \rangle}(q^2)$ represent the terms that can be directly transformed by the Borel transformation without the need for a dispersion relation. By using the Borel transformation to Eq.~\eqref{Pi-OPE}, we have
\begin{eqnarray}\label{Pi-MB}
  \Pi^{\text{OPE}}(M_B^2) = \int_{(4 m_Q)^2}^\infty ds \rho^{\text{OPE}}(s) e^{-s/M_B^2} + \Pi^{\langle GG \rangle}(M_B^2) + \Pi^{\langle GGG \rangle}(M_B^2).
\end{eqnarray}

\begin{figure}[ht]
  \centering
  \includegraphics[width=12cm]{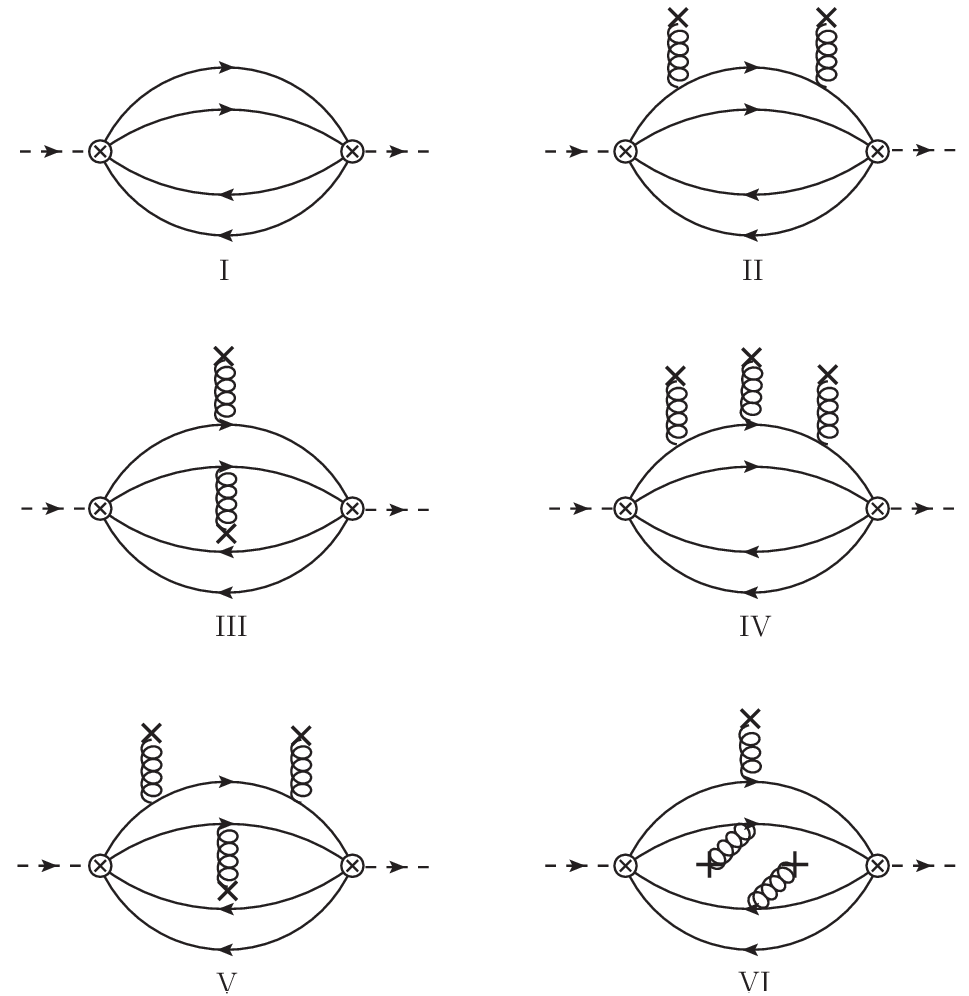}
  \caption{The standard leading-order (LO) Feynman diagrams for a fully-charm tetraquark state, which contribute to the spectral densities as seen in Eq.~\eqref{Pi-MB}, implicitly include permutation diagrams. Diagram I illustrates the perturbative contribution, while diagrams II, III represent the contributions from two-gluon condensates, and diagrams IV-VI are three-gluon condensate contributions. All diagrams up to the three-gluon condensate at the LO of $\alpha_s$ are showcased.}\label{Feyn-Diag}
\end{figure}

The typical LO Feynman diagrams of a fully-charm tetraquark state that contribute to the Eq.~\eqref{Pi-MB} are shown in Fig.~\ref{Feyn-Diag}, where diagram I represents the contribution from perturbative part, diagrams II and III denote the two-gluon condensate, diagrams IV-VI are the trigluon condensates. As an example, the lengthy expressions $\rho^{\text{OPE}}(s)$, $\Pi^{\langle GG \rangle}(M_B^2)$, and $\Pi^{\langle GGG \rangle}(M_B^2)$ in Eq.~\eqref{Pi-MB} for $0^{-+}$ tetraquark states considered in this study will be included in the Appendix.

From the phenomenological perspective, by segregating the ground state contribution of the tetraquark state, the correlation function $\Pi(q^{2})$ can be expressed as a dispersion integral across the physical spectrum,
\begin{eqnarray}\label{Pi-hadron}
  \Pi(q^2) = \frac{(\lambda_X)^2}{(M_X)^2 - q^2} + \frac{1}{\pi} \int_{s_0}^\infty ds \frac{\rho^h(s)}{s - q^2}.
\end{eqnarray}
Here, the superscript X signifies the lowest-lying tetraquark state, with $M_{X}$ representing its mass, and $\rho^h(s)$ being the spectral density. This density includes contributions from both higher excited states and continuum states beyond the threshold $s_{0}$. Following the methodology in Refs.~\cite{Reinders:1984sr,Colangelo:2000dp}, for calculating the phenomenological aspect of QCD sum rules, a comprehensive series of intermediate states must be inserted amidst two color octet-octet tetraquark interpolating currents. This summation encompasses all potential hadronic states elicited by the color octet-octet tetraquark current. The coupling constant $\lambda_X$ is defined as Ref.~\cite{Wang:2022xja}:
\begin{eqnarray}
\langle 0|j^{0^{-+}}(0)|X\rangle &=& \lambda_X.
\end{eqnarray}
By applying a Borel transformation to the phenomenological side as indicated in Eq.~\eqref{Pi-hadron} and aligning it with Eq.~\eqref{Pi-MB}, we obtain the master equation
\begin{eqnarray}\label{main-equation}
(\lambda_{X})^{2}\exp\left(-\frac{M_{X}^{2}}{M_{B}^{2}}\right)=\int_{(4 m_Q)^2}^{s_{0}} ds \rho^{\text{OPE}}(s) e^{-s/M_B^2} + \Pi^{\langle GG \rangle}(M_B^2) + \Pi^{\langle GGG \rangle}(M_B^2).
\end{eqnarray}
Finally, we differentiate Eq.~\eqref{main-equation} with respect to $\frac{1}{M_{B}^{2}}$ and obtain the mass of the tetraquark state
\begin{eqnarray}\label{main-function}
  M_X(s_0, M_B^2) &=& \sqrt{-\frac{L_1(s_0, M_B^2)}{L_0(s_0, M_B^2)}},
\end{eqnarray}
where the moments $L_{0}$ and $L_{1}$ are, respectively, defined as
\begin{eqnarray}\label{OPE-function}
  L_0(s_0, M_B^2) &=& \int_{(4 m_Q)^2}^{s_{0}} ds \rho^{\text{OPE}}(s) e^{-s/M_B^2} + \Pi^{\langle GG \rangle}(M_B^2) + \Pi^{\langle GGG \rangle}(M_B^2), \label{L0} \\
  L_1(s_0, M_B^2) &=& \frac{\partial}{\partial (M_B^2)^{-1}} L_0(s_0, M_B^2).
\end{eqnarray}

\section{Numerical analysis}
For conducting the numerical analysis of the QCD sum rules, we employ specific values for the masses of heavy quarks, the strong coupling, and the QCD condensates. It has been determined in~\cite{Shifman:1978bx, Shifman:1978by, Reinders:1984sr, Narison:1989aq, Colangelo:2000dp}, for numerical analyses, we take $m_c (m_c) = \overline{m}_c= (1.27 \pm 0.03) \; \text{GeV}$, where the ``running mass" is used for the charm quark in the $\overline{\text{MS}}$ scheme. For two-gluon and three-gluon condensates, we take the values: $\langle\alpha_{s}G^{2}\rangle=(6.35\pm0.35)\times 10^{-2} \; \text{GeV}^{4}$, $\langle g_{s}^{3}G^{3}\rangle=(8.2\pm1.0)\times \langle\alpha_{s}G^{2}\rangle \; \text{GeV}^{2}$.

In QCD sum rules, two crucial parameters are introduced: $M_{B}^{2}$ and $s_{0}$, which correspond to the threshold parameter and the Borel parameter, respectively. With a specific value of \(s_0\) set, the Borel parameter \(M^2_B\) is determined based on three essential conditions~\cite{Shifman:1978bx,Shifman:1978by,Reinders:1984sr,Colangelo:2000dp}. Primarily, to accurately isolate the ground state characteristics of the tetraquark state, it is imperative to ensure that the contribution from the continuum is subordinate to that of the pole contribution (PC)~\cite{Colangelo:2000dp,Matheus:2006xi}. This condition can be quantified using a specific formula.
\begin{eqnarray}
  R_{i}^{\text{PC}}(s_0, M_B^2) = \frac{L_0(s_0, M_B^2)}{L_0(\infty, M_B^2)} \; , \label{RatioPC}
\end{eqnarray}
where the subscript $i$ encompasses all currents. With this precondition, the effects of higher excited and continuum states are diminished. This requirement leads to the establishment of a pivotal $M_{B}^{2}$ value, defined as the upper threshold $(M_{B}^{2})_{max}$. The OPE's convergence sets the lower threshold of $M_{B}^{2}$, dubbed $(M_{B}^{2})_{min}$, acting as the secondary guideline. Generally, the level of convergence is quantified by comparing the ratio of condensate contributions to the total contributions.
\begin{eqnarray}
  R_{i}^{\text{cond}}(s_0, M_B^2) = \frac{L_0^{\text{dim}}(s_0, M_B^2)}{L_0(s_0, M_B^2)}\, ,
\end{eqnarray}
where the superscript ``dim" represents the dimension of the relevant condensate in the Operator Product Expansion (OPE), as described in Eq.~\eqref{OPE-function}. Therefore, for a given $s_{0}$, the suitable Borel window, which is the range between $(M_{B}^{2})_{min}$ and $(M_{B}^{2})_{max}$, has been determined.

In practical application, to verify whether the OPE convergence criterion is achieved, we first ensure that the contribution from the highest-order condensate, $\langle G^{3}\rangle$, does not surpass 10\% of the total OPE contribution in this work. The third condition requires that the tetraquark state mass $M_{X}$ shows minimal dependence on the parameter $s_{0}$. To determine the appropriate continuum threshold $s_{0}$ accurately, we performed an analysis using methods similar to those described in Refs.~\cite{Finazzo:2011he,Qiao:2013raa,Qiao:2013dda}. It's crucial to recognize that $s_{0}$ is linked to the ground state mass as $\sqrt{s_{0}} \approx (M_{X} + \delta)$ GeV, with $\delta$ varying between 0.40 and 0.80 GeV. Therefore, in the numerical evaluation, exploring different $\sqrt{s_{0}}$ values is essential to meet this criterion. From these options, we need to choose a value that sets the most appropriate Borel window for the parameter $M_{B}^{2}$. Ideally, within this optimal window, the mass $M_{X}$ of the fully-charm tetraquark should exhibit minimal variation with the Borel parameter $M_{B}^{2}$. The selected $\sqrt{s_{0}}$ value, which corresponds to the most favorable mass curve, will serve as the central value. To accommodate the uncertainty in $s_{0}$, we adjust $\sqrt{s_{0}}$ by 0.20 GeV in our calculations, thus establishing the upper and lower bounds for $\sqrt{s_{0}}$.
\begin{figure}[htb]
\begin{center}
\includegraphics[width=7.6cm]{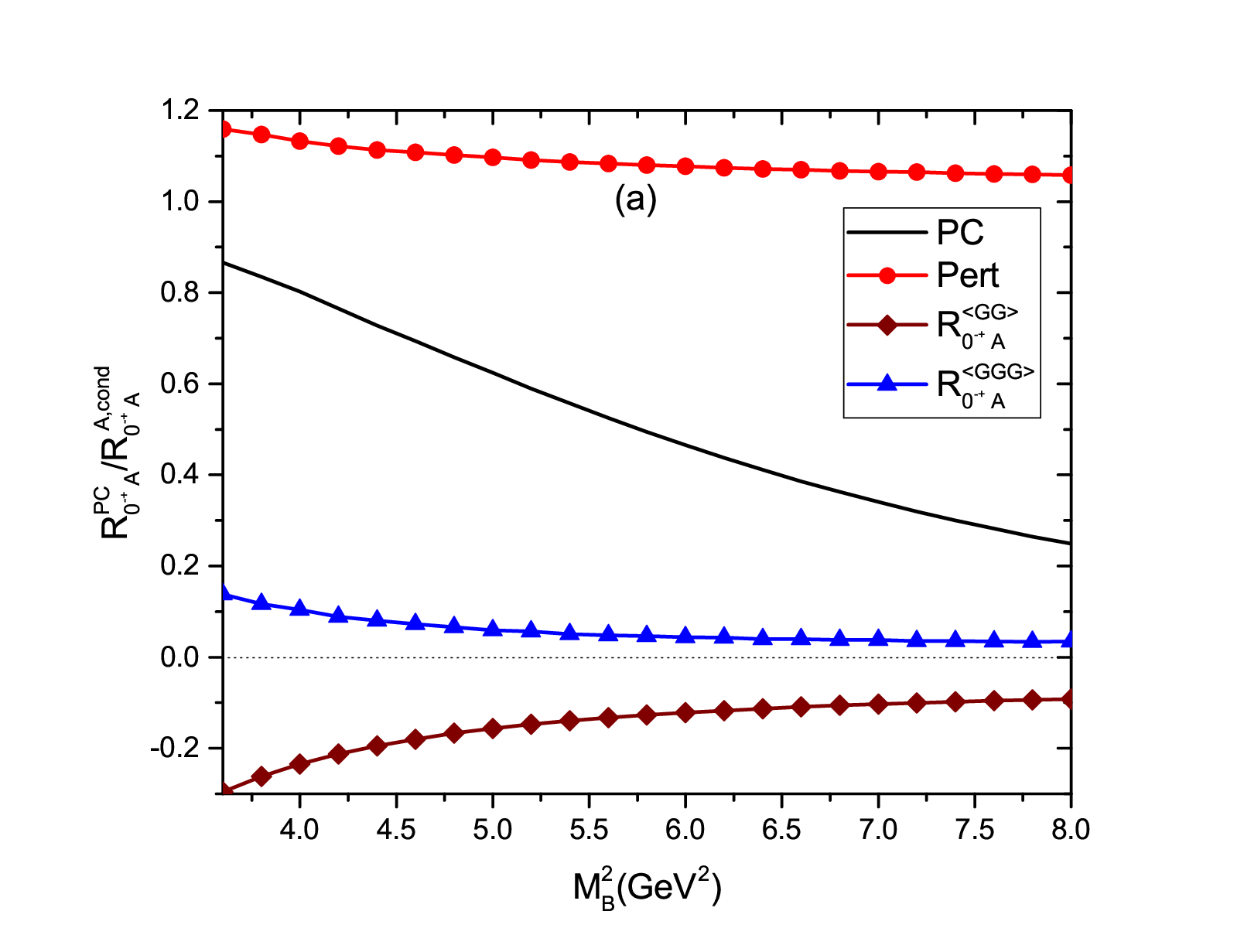}
\includegraphics[width=7.6cm]{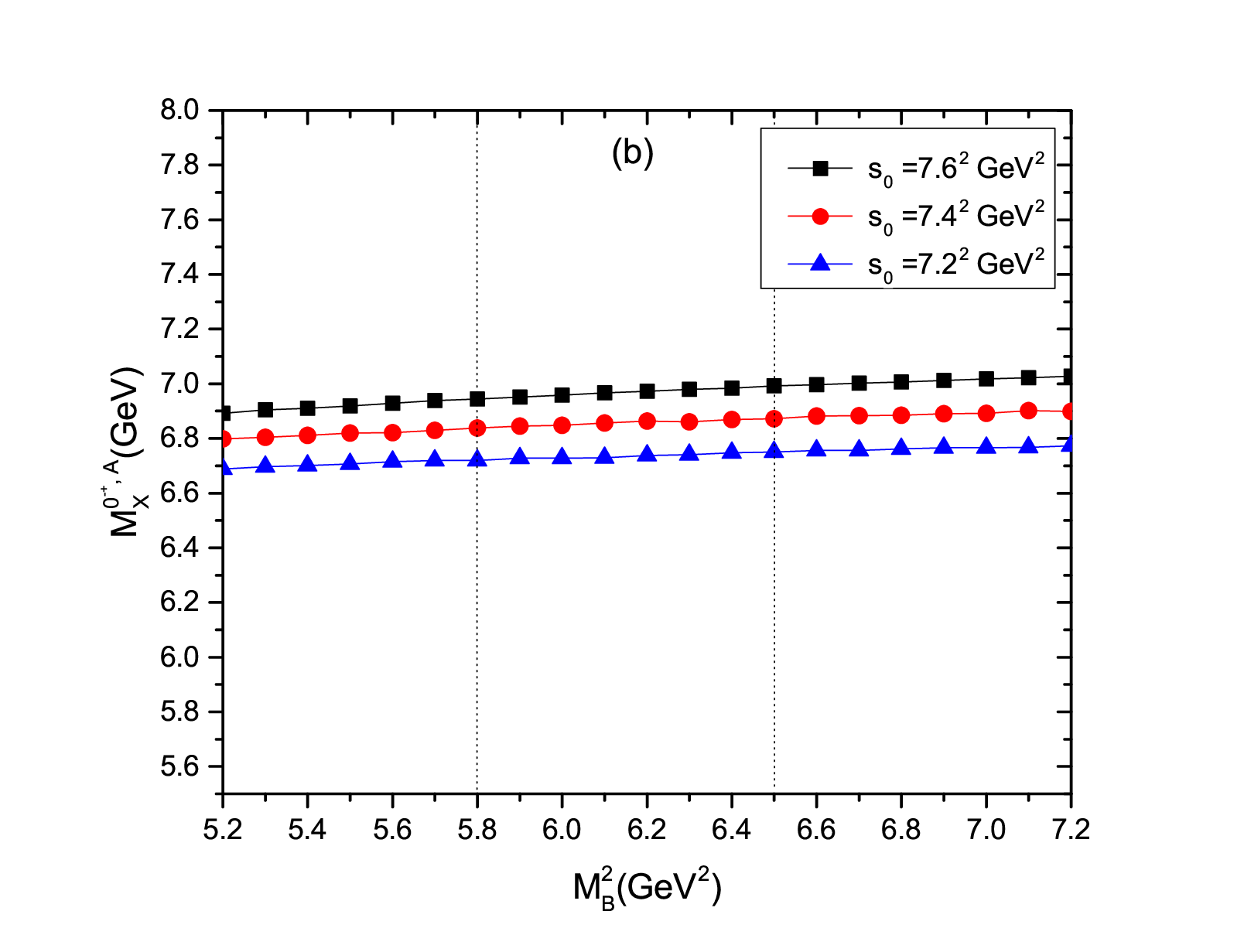}
\caption{(color). The figures for current $j_{A}^{0^{-+}}$. (a) The pole contribution ratio $R_{0^{-+}}^{\text{PC}}$ and OPE convergence ratio $R_{0^{-+}}^{\text{cond}}$ as functions of the Borel parameter $M_B^2$ with the central value of $s_{0}$; (b) The masses $M_{X}^{0^{-+}}$ as functions of $M_B^2$ for $s_0= 7.20^{2}\, \text{GeV}^{2}$, $7.40^{2}\, \text{GeV}^{2}$, and $7.60^{2} \, \text{GeV}^{2}$ from down to up, respectively, and the two vertical lines indicate the upper and lower bounds of valid Borel window with the central value of $s_{0}$.}
\label{fig2}
\end{center}
\end{figure}
\begin{figure}[htb]
\begin{center}
\includegraphics[width=7.6cm]{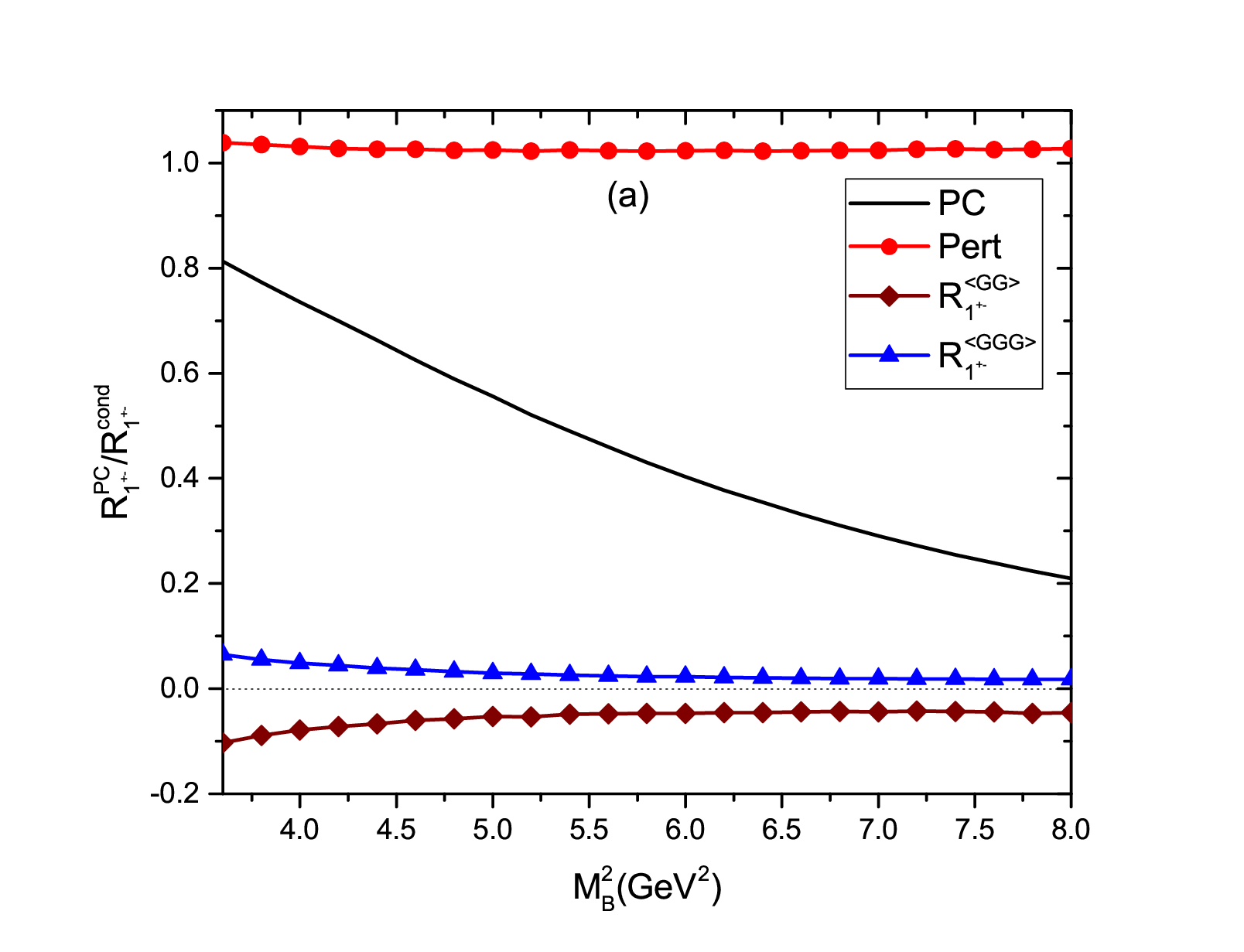}
\includegraphics[width=7.6cm]{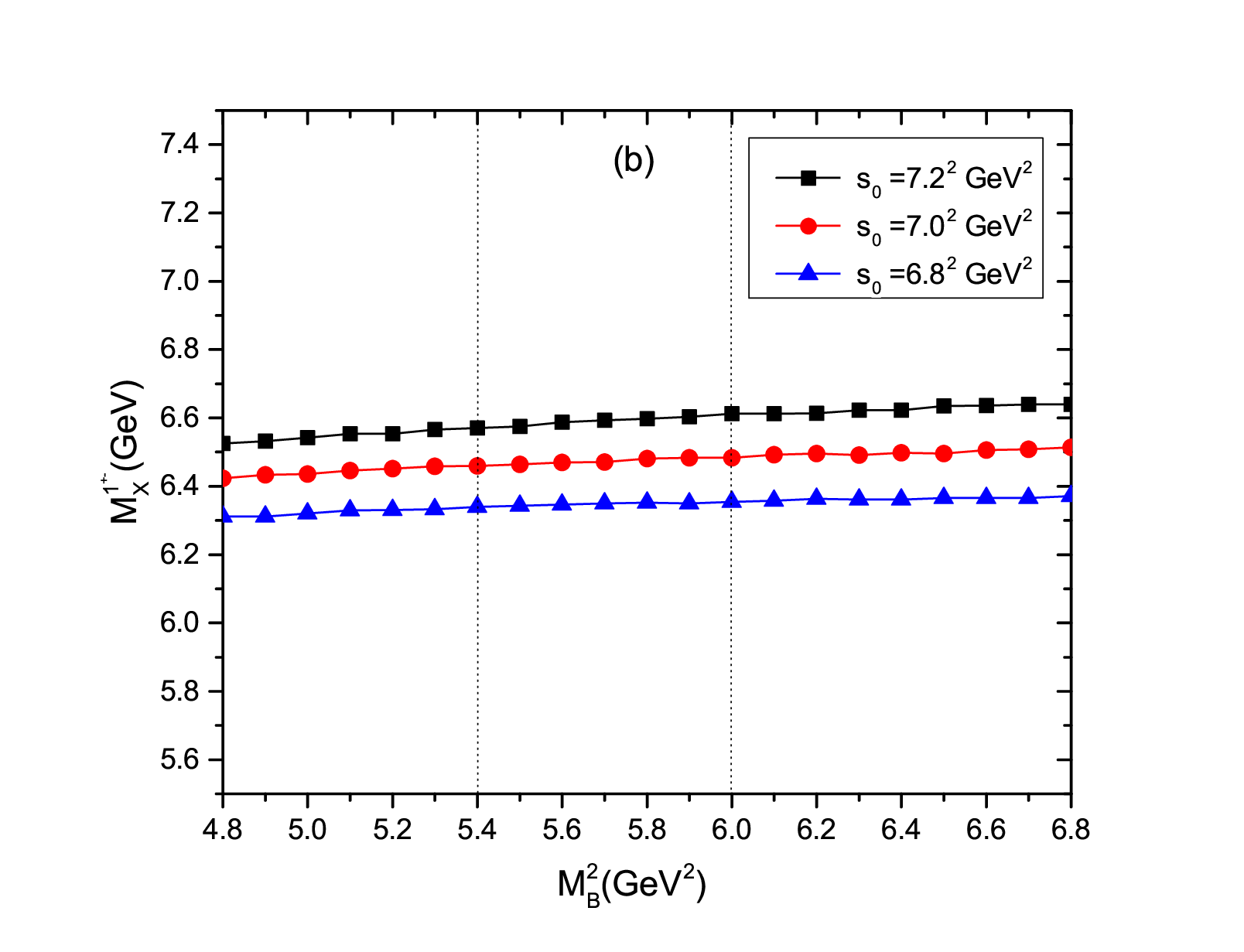}
\caption{(color). The same caption as in Fig.~\ref{fig2}, but for current $j_{\mu}^{1^{+-}}$.}
\label{fig3}
\end{center}
\end{figure}
\begin{figure}[htb]
\begin{center}
\includegraphics[width=7.6cm]{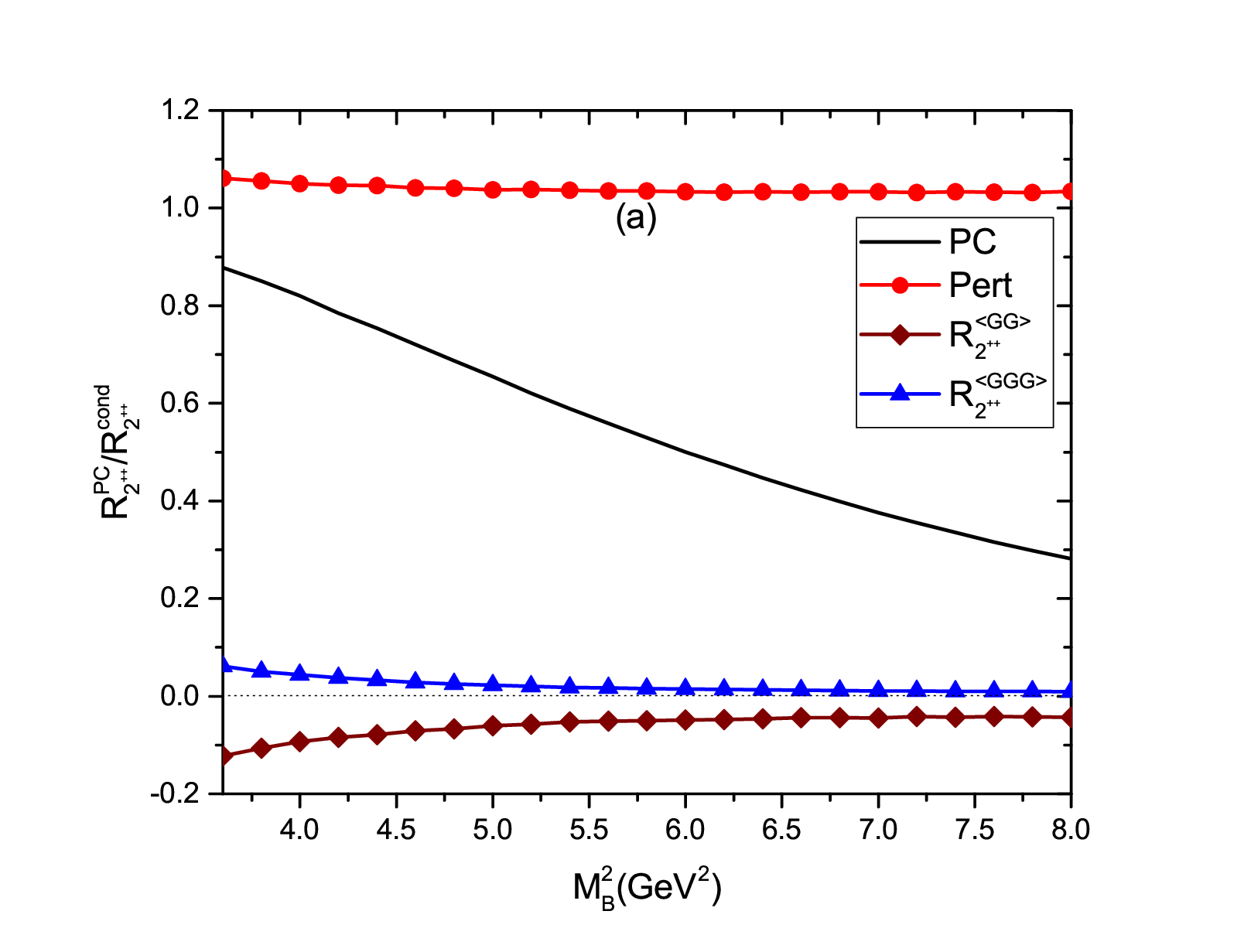}
\includegraphics[width=7.6cm]{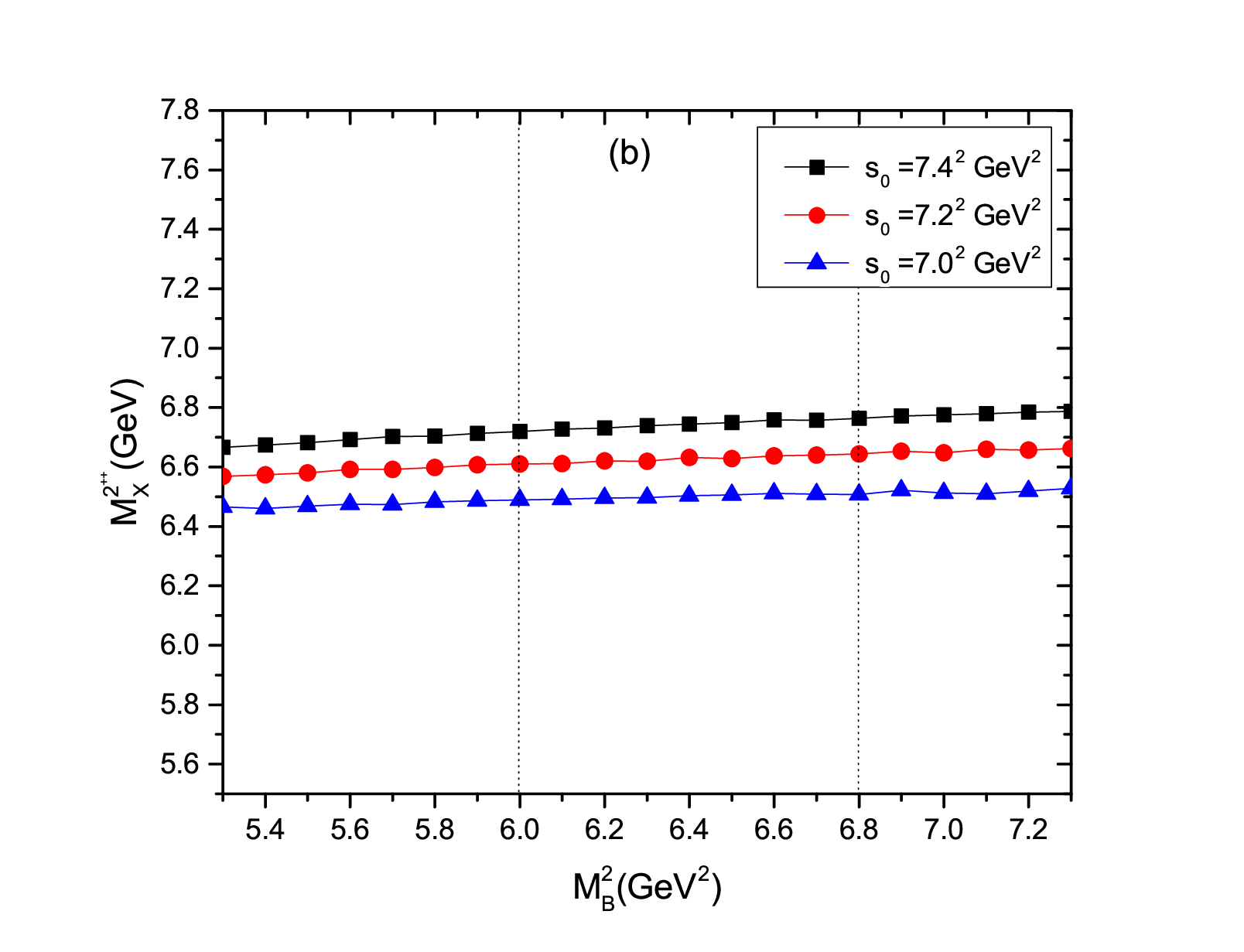}
\caption{(color). The same caption as in Fig.~\ref{fig2}, but for current $j_{\mu\nu}^{2^{++}}$.}
\label{fig4}
\end{center}
\end{figure}

For the fully-charm tetraquark state characterized by $J^{PC} = 0^{-+}$, derived from the current $j_{A}^{0^{-+}}(x)$, we present two ratios, $R^{\text{PC}}_{0^{-+},A}$ and $R^{\text{cond}}_{0^{-+},A}$, as functions of the Borel parameter $M_{B}^{2}$ in Fig.~\ref{fig2}(a) at the chosen $s_{0} = 7.40^{2}\, \text{GeV}^{2}$, and depict the mass as a function of $M_{B}^{2}$ for varying $s_{0}$ values in Fig.~\ref{fig2}(b). Fig.~\ref{fig2}(b) also includes two lines indicating the upper and lower limits of the valid Borel window at the central $s_{0}$ value. The selection of the Borel parameter $M_{B}^{2}$ and the continuum threshold $s_{0}$ is guided by two criteria: ensuring pole dominance in the phenomenological aspect and securing OPE on the QCD side, in line with QCD sum rule principles. Additionally, we apply the rule $\sqrt{s_{0}} = M_{X} + (0.40-0.80) \, \text{GeV}$ as a further guideline. Similar analyses apply to the fully-charm tetraquark states with $J^{PC} = 1^{+-}$ from the current $j_{\mu}^{1^{+-}}$ and $J^{PC} = 2^{++}$ from the current $j_{\mu\nu}^{2^{++}}$, depicted in Fig.~\ref{fig3} and Fig.~\ref{fig4}, respectively.

For simplicity, the article includes only representative figures for currents with quantum numbers $J^{PC} = 0^{-+}$, $1^{+-}$, and $2^{++}$, omitting figures for other cases.

The derived Borel parameters, continuum threshold values, pole contributions, masses, and pole residues are concisely tabulated in Table \ref{tab1}. This table clearly demonstrates that pole dominance on the phenomenological side is adequately achieved. Moreover, within these Borel windows, the mass curves present a stable platform. Adhering to the three foundational principles of QCD sum rules, we offer plausible mass estimations as collected in Table \ref{tab1}. The central values are in alignment with the most stable outcomes regarding $M_{B}^{2}$, while the margin of error stems from the variability in condensates, quark mass, and the parameters $s_{0}$ and $M_{B}^{2}$.

\begin{table}[htb]
\begin{center}
\begin{tabular}{|c|c|c|c|c|c|c|c|c}\hline\hline
 $J^{PC}$& $M_B^2\,(\rm{GeV}^2)$  & $\sqrt{s_{0}}\,(\rm{GeV})$ & PC & $M_{X}$ $(\rm{GeV})$ & $\lambda_{X}$ $(\rm{10^{-1}GeV^{5}} )$\\ \hline
 $j_{A}^{0^{-+}}$ & $5.80\!-\!6.50$ & $7.40\pm 0.20$  & $(49\!-\!40)\%$ & $6.86_{-0.14}^{+0.14}$  & $0.71^{+0.08}_{-0.08}$  \\ \hline
 $j_{B}^{0^{-+}}$ & $5.80\!-\!6.50$ & $7.40\pm 0.20$  & $(50\!-\!40)\%$ & $6.85_{-0.15}^{+0.13}$  & $5.00^{+0.55}_{-0.58}$  \\ \hline
 $j^{0^{--}}$ & $6.50\!-\!7.20$ & $7.60\pm 0.20$  & $(51\!-\!40)\%$ & $7.00_{-0.14}^{+0.13}$   & $1.97^{+0.20}_{-0.21}$ \\ \hline
 $j_{A\mu}^{1^{--}}$ & $6.40\!-\!7.20$ & $7.60\pm 0.20$  & $(51\!-\!40)\%$ & $7.01_{-0.14}^{+0.14}$   & $0.91^{+0.10}_{-0.10}$ \\ \hline
 $j_{B\mu}^{1^{--}}$ & $6.50\!-\!7.20$ & $7.60\pm 0.20$  & $(50\!-\!40)\%$ & $7.00_{-0.13}^{+0.14}$   & $1.55^{+0.16}_{-0.17}$ \\ \hline
 $j_{\mu}^{1^{+-}}$ & $5.40\!-\!6.00$ & $7.00\pm 0.20$ & $(49\!-\!40)\%$  & $6.48_{-0.14}^{+0.12}$    & $0.74^{+0.08}_{-0.09}$  \\ \hline
 $j_{A\mu}^{1^{-+}}$ & $6.40\!-\!7.30$ & $7.60\pm 0.20$  & $(49\!-\!40)\%$ & $7.02_{-0.14}^{+0.14}$   & $0.80^{+0.08}_{-0.09}$ \\ \hline
$j_{B\mu}^{1^{-+}}$ & $6.30\!-\!7.30$ & $7.60\pm 0.20$ & $(50\!-\!40)\%$ & $7.01_{-0.14}^{+0.15}$    & $1.51^{+0.15}_{-0.15}$\\ \hline
$j_{\mu\nu}^{2^{++}}$ & $6.00\!-\!6.80$ & $7.20\pm 0.20$  & $(50\!-\!40)\%$ & $6.62_{-0.13}^{+0.14}$    & $3.20^{+0.31}_{-0.34}$\\ \hline
 \hline
\end{tabular}
\end{center}
\caption{The ranges of Borel parameter $M_B^2$, threshold parameter $s_0$, pole contributions (PC), the masses, and pole residues of ground states for the nine fully-charm tetraquark states considered in this work.}
\label{tab1}
\end{table}

\section{Conclusions}
In this work, we conduct a comprehensive analysis of the mass spectra for fully-charm tetraquark states in the $8_{[c\bar{c}]}\otimes8_{[c\bar{c}]}$ color configuration through QCD sum rules. By utilizing currents of type $8_{[c\bar{c}]}\otimes8_{[c\bar{c}]}$ with quantum numbers $J^{PC}=0^{-+},0^{--},1^{-+},1^{+-},1^{--}$, and $2^{++}$, respectively shown in Eqs.~(\ref{current-1}-\ref{current-9}), we carry out analytical calculations up to dimension six in OPE. Then, we perform numerical analyses of these nine cases and summarize the masses of the tetraquark states in $8_{[c\bar{c}]}\otimes8_{[c\bar{c}]}$ configurations in Table \ref{tab1}.

The findings indicate that the fully-charm tetraquark states with quantum numbers $J^{PC}=1^{+-}$ and $2^{++}$ have masses ranging from 6.48 to 6.62 GeV. Meanwhile, those with $J^{PC}=0^{-+},0^{--},1^{--},1^{-+}$ are found to be in the range of 6.85 to 7.02 GeV. Significantly, the mass estimates for the $c\bar{c}c\bar{c}$ tetraquarks, especially those with $J^{PC}=2^{++}$, correspond with the broad structure detected by LHCb. Additionally, the predicted masses for fully-charm tetraquarks with $J^{PC}=0^{-+}$ and $1^{-+}$ are expected to be very close to the mass of X(6900), within the error margins. Such results suggest the possible presence of some $8_{[c\bar{c}]}\otimes8_{[c\bar{c}]}$ components in the di-$J/\psi$ structures observed by LHCb, CMS, and ATLAS Collaborations.

%%%%%%%%%%%%%%%%%%%%%%%%%%%%%%%%%%%%%%%%%%%%%%%%%%%%%%%%%%%%%%%%%%%%%%
\vspace{.7cm} {\bf Acknowledgments} \vspace{.3cm}

This work is supported by the National Natural Science Foundation of China (NSFC) under Grant No. 11975090 and the Natural Science Foundation of Hebei Province under Grant No. A2023205038.

%%%%%%%%%%%%%%%%%%%%%%%%%%%%%%%%%%%%%%%%%%%%%%%%%%%%%%%%%%%%%%%%%%%%%%%

\begin{widetext}

  \newpage

\appendix \label{appendix}

\textbf{Appendix}

In this appendix, as an example, we present the spectral density $\rho^{\text{OPE}}(s)$ in Eq.~(\ref{Pi-OPE}) for current $j^{0^{-+}}_{A}$ shown in Eq.~(\ref{current-1}) as follows:
\begin{eqnarray}
\rho_{0^{-+},\,A}^{\text{pert}}(s) &=& \frac{1}{2\times  \pi^{6}}\int_{x_{i}}^{x_{f}} dx\int_{y_{i}}^{y_{f}} dy\int_{z_{i}}^{z_{f}}dz \left\{ \frac{F_{xyz}^{4}A_{xyz}xyz}{256} \right.\nonumber\\ &-& \frac{F_{xyz}^{3}(2A_{xyz}m_{Q}^{2}z+6A_{xyz}sxyz-2m_{Q}^{2}xy)} {384}\nonumber\\&+& \left.\frac{F_{xyz}^{2} (A_{xyz}m_{Q}^{2}sz + A_{xyz}s^{2}xyz - m_{Q}^{4}-m_{Q}^{2}sxy)}{128}\right\},
\end{eqnarray}
\begin{eqnarray}
\rho_{0^{-+},\,A}^{\langle GG\rangle}(s) &=& -\frac{\langle g_{s}^{2}GG\rangle}{2^{13}\times9\pi^{6}} \int_{x_{i}}^{x_{f}} dx\int_{y_{i}}^{y_{f}} dy \int_{z_{i}}^{z_{f}} dz\frac{1}{A_{xyz}^{3}x^{3}y^{3}z^{3}}\nonumber\\
&\times&\left\{ A_{xyz}^{4}z \big[9F_{xyz}^{2}x^{3}y^{3}z^{3}-12F_{xyz}m_{Q}^{2}xy(4x^{3}(y^{3}+z^{3}) - 4x^{2}z^{3}+xy z^{3} \right.\nonumber\\ &+& \left.4(y-1)y^{2}z^{3})+16m_{Q}^{4}(x^{3}(y^{3}+z^{3})+y^{3}z^{3})\big]\right.
\nonumber\\
&+& \left. A_{xyz}^{3}xy\big[9F_{xyz}^{2}x^{3}y^{3}z^{3}-48F_{xyz}m_{Q}^{2}x^{3}y^{3}z - 2m_{Q}^{4}(8x^{3}(y^{3}+z^{3})\right.\nonumber\\
&-&\left.12x^{2}z(y^{2}+z^{2})+3xyz^{3}+4y^{2}(2y-3)z^{3}) \big] \right.\nonumber\\
&+& \left. 3A_{xyz}^{3}s^{2}x^{3}y^{3}z^{3}(A_{xyz}z+xy) - 6A_{xyz}^{2}m_{Q}^{2}x^{3}y^{3}z^{2}(m_{Q}^{2}-2F_{xyz}xy)\right.\nonumber\\
&+& \left. 6A_{xyz}sxyz\big[ A_{xyz}^{3}(m_{Q}^{2}(8x^{3}(y^{3}+z^{3})-4x^{2}z^{3}+xyz^{3}+4y^{2}(2y-1)z^{3}) \right.\nonumber\\
&-& \left.3F_{xyz}x^{2}y^{2}z^{3})+A_{xyz}^{2}x^{3}y^{3}(4m_{Q}^{2}-3F_{xyz}z^{2}) - A_{xyz}m_{Q}^{2}x^{3}y^{3}z \right.\nonumber\\
&+&\left. 4m_{Q}^{2}x^{3}y^{3}z^{2}(2z+1) + 8A_{xyz}m_{Q}^{2}x^{3}y^{3}z^{3}(m_{Q}^{2}(2z+3)-6F_{xyz}xy(z+1))\right.\nonumber\\ &-&\left. 16m_{Q}^{4}x^{4}y^{4}z^{3}\big]\right\},
\end{eqnarray}

\begin{eqnarray}
  \Pi^{\langle GG\rangle}_{0^{-+},\,A}(M_B^2)&=& -\frac{\langle g_{s}^{2}GG\rangle}{2^{10}\times 9\pi^{6}} \int_{0}^{1} dx \int_{0}^{1-x} dy\int_{0}^{1-x-y}dz \frac{e^{-\frac{f_{xyz}m_{Q}^{2}}{M_B^2}}}{x^{3}y^{3}z^{3}A_{xyz}^{3}}\nonumber\\
&\times& \left\{ m_{Q}^{6}\big[ 1+f_{xyz}z(-1+y+z)+f_{xyz}x(y+z) - f_{xyz}^{2}xyzA_{xyz}\big]\right.\nonumber\\
&\times&\big[ 3x^{2}y^{3}z^{3}(-1+y+z)
 + 3xy^{3}z^{3}(-1+y+z)^{2} + y^{3}z^{3}(-1+y+z)^{3} \nonumber\\
&+& 3x^{4}(-1+y+z)^{2}(y^{3}+z^{3})
  + x^{6}(y^{3}+z^{3}) + x^{3}(-1+y+z)^{3}(y^{3}+z^{3}) \nonumber\\
&+& \left. 3x^{5}(y^{4}+y^{3}(z-1)+yz^{3}
  + (z-1)z^{3})\big]\right\},
\end{eqnarray}

\begin{eqnarray}
\rho_{0^{-+},\,A}^{\langle GGG \rangle }(s) &=& \frac{\langle g_{s}^{3}GGG\rangle }{2^{14}\times9\pi^{6}} \int_{x_{i}}^{x_{f}} dx\int_{y_{i}}^{y_{f}} dy\int_{z_{i}}^{z_{f}} dz \frac{1}{A_{xyz}^{3}x^{3}y^{3}z^{3}}\left\{ A_{xyz}^{4}\big[3xyz\right.\nonumber\\
&\times&\left.(F_{xyz}-s)(8x^{3}(y^{3}+z^{3})+3x^{2}yz^{3} +3xy^{2}z^{3} + 8y^{3}z^{3})+2m_{Q}^{2}(24x^{4}(y^{4}+z^{4})\right.\nonumber\\
&+&\left.x^{3}(-4y^{3}z+3yz^{4} -24z^{4})- 6x^{2}yz^{4}+3x(y-2)y^{2}z^{4} + 24(y-1)y^{3}z^{4})) \right.\nonumber\\
&+&\left. A_{xyz}^{3}(9x^{4}y^{4}z^{2}(F_{xyz}-s) + m_{Q}^{2}xy(x^{3}(8z^{3}+6y^{3}(z+8)) -3x^{2}yz^{2}(y-z) \right.\nonumber\\
&+&\left. 3xy^{2}z^{3} + 8y^{3}z^{3})) + 3A_{xyz}^{2}x^{3}y^{3}z(3xyz^{2}(F_{xyz}-s) - m_{Q}^{2}(z^{2}-4xy)) \right.\nonumber\\
&+&\left. 2A_{xyz}x^{3}y^{3}z^{2}(12xyz^{2}(F_{xyz}-s) + m_{Q}^{2}(3xy(z+2)-4z^{2})) \right.\nonumber\\
&+&\left. 48m_{Q}^{2}x^{4}y^{4}z^{3}(z+1)\big]\right\},
\end{eqnarray}

\begin{eqnarray}
  \Pi^{\langle GGG \rangle}_{0^{-+},\,A}(M_B^2)&=&-\frac{\langle g_{s}^{3}GGG\rangle}{2^{15}\times 9\pi^{6}} \int_{0}^{1} dx \int_{0}^{1-x} dy\int_{0}^{1-x-y}dz\nonumber\\
&\times&\frac{e^{-\frac{f_{xyz}m_{Q}^{2}}{M_B^2}}}{x^{4}y^{4}z^{4}A_{xyz}^{4}}  \left\{m_{Q}^{4}\big[ (2(8f_{xyz}^{2}yz(y^{4}+z^{4})x^{6}+ f_{xyz}(8f_{xyz}zy^{6}+(f_{xyz}z(7z-8) \right.\nonumber\\
&+&8zy^{4} + 9f_{xyz}z^{5}y^{2} + 8z^{4}(f_{xyz}(z-1)z+1)y+8z^{5})x^{5}+8)y^{5}+ (8f_{xyz}zy^{5}\nonumber\\
&+&\left. (f_{xyz}z(7z-8)+8)y^{4} - f_{xyz}^{2}z^{5}y^{3} - f_{xyz}z^{4}(f_{xyz}(z-1)z+1)y^{2}\right.\nonumber\\
&+&\left. 9f_{xyz}z^{5}y + 8z^{4}(f_{xyz}(z-1)z+1))x^{4} + yz^{4}(f_{xyz}^{2}zy^{3} + f_{xyz}z(y+z-1)\right.\nonumber\\
 &+&\left.1)x^{3}
  + f_{xyz}y^{3}z^{4}(9f_{xyz}zy^{2} + (f_{xyz}(z-1)zy+y+z)x^{2}\right.\nonumber\\
   &+&\left. y^{3}z^{4}(8f_{xyz}^{2}z(y+z-1)y^{2}
  +f_{xyz}(8y^{2} + 9zy+(z-1)z)+1)x \right.\nonumber\\
  &+&\left. 8y^{4}z^{4}(f_{xyz}z(y+z-1)+1))m_{Q}^{2}
  + M_{B}^{2}(-8f_{xyz}yz((f_{xyz}z-8)y^{4}\right.\nonumber\\
   &+&\left. f_{xyz}z^{4}y - 8z^{4})x^{6} + (-8f_{xyz}z(f_{xyz}z-8)y^{6}
   + (f_{xyz}^{2}(8-5z)z^{2}\right.\nonumber\\
    &+& \left.56f_{xyz}(z-2)z+16)y^{5} - 8z(f_{xyz}z-2)y^{4} - 11f_{xyz}^{2}z^{5}y^{3}\right.\nonumber\\
    &-&\left. 8f_{xyz}z^{4}(f_{xyz}z^{2}-(f_{xyz}+9)z+1)y^{2} + 16z^{4}(f_{xyz}z(4z-7)+1)y\right.\nonumber\\
    &+&\left.16z^{5})x^{5}+ z(-8(f_{xyz}z-2)y^{5} + (-6f_{xyz}^{2}z^{4} - 5f_{xyz}z^{2} + 2(4f_{xyz}+7)z\right.\nonumber\\
    &-&\left.64)y^{4} + f_{xyz}z^{3}(-3f_{xyz}z^{2} + (3f_{xyz} + 8)z+3)y^{3} + 2z^{3}(2f_{xyz}z(2z-17)\right.\nonumber\\
    &+&\left.1)y^{2} - 6z^{3}(8f_{xyz}z^{2} - (8f_{xyz}+3)z+8)y + 16(z-1)z^{4})x^{4}\right.\nonumber\\
    &-&\left. yz^{4}(11f_{xyz}^{2}zy^{4} +f_{xyz}(3f_{xyz}z^{2} - (3f_{xyz} + 8)z+3)y^{3} + 24f_{xyz}zy^{2} \right.\nonumber\\
    &+&\left. 2(6f_{xyz}z^{2}-(6f_{xyz} + 1)z+6)y - 2(z-1)z)x^{3} - 2y^{3}z^{4}(4f_{xyz}^{2}zy^{3} \right.\nonumber\\
    &+&\left. 4f_{xyz}(f_{xyz}z^{2}- (f_{xyz}+9)z+1)y^{2} + (2f_{xyz}z(17-2z)-1)y +6f_{xyz}z^{2}\right.\nonumber\\
     &-&\left. 6f_{xyz}z - z+6)x^{2}+ 2y^{3}z^{4}(4f_{xyz}zy^{3} + 8(f_{xyz}z(4z-7)+1)y^{2} \right.\nonumber\\
     &-&\left. 3(8f_{xyz}z^{2} - (8f_{xyz}+3)z+8)y+(z-1)z)x + 16y^{4}z^{5}(y+z\right.\nonumber\\
     &-&\left.1)))A_{xyz}^{4} + x^{4}y^{4}z(2(f_{xyz}xy+1)m_{Q}^{2}
       + 3M_{B}^{2}z(f_{xyz}z^{2}-4)\right.\nonumber\\
       &+&\left.M_{B}^{2}xy(3f_{xyz}^{2}z^{3}-12f_{xyz}z+2)) A_{xyz}^{3}
       +2x^{4}y^{4}z^{3}(M_{B}^{2}(4f_{xyz}^{2}xyz^{2}\right.\nonumber\\
       &-&\left.z+2f_{xyz}xy(z-3)-6)-f_{xyz}mQ^{2}(f_{xyz}xy+1)z)A_{xyz}^{2} \right.\nonumber\\
       &+&\left.2x^{4}y^{4}z^{3}(M_{B}^{2}(6f_{xyz}yzx^{2}+y(2f_{xyz}(3y-13z -15)z+1)x\right.\nonumber\\
       &-&\left.8z(z+3))-mQ^{2}(f_{xyz}xy+1)(8f_{xyz}z^{2}-1))A_{xyz}\right.\nonumber\\
       &+&\left. 16x^{4}y^{4}((f_{xyz}xy+1)mQ^{2}+M_{B}^{2}xy)z^{4}
       \right\},
\end{eqnarray}

where we have used the following definitions:
\begin{eqnarray}
 A_{x} &=& 1-x,\;A_{xy}=1-x-y,\;A_{xyz}=1-x-y-z, \\
 F_{xyz} &=& m_{Q}^{2}f_{xyz} - s, \\
 f_{xyz} &=& \frac{1}{x}+\frac{1}{y}+\frac{1}{z}+\frac{1}{1-x-y-z}, \\
 x_{f/i} &=& \frac{1}{2} \left(1\pm \sqrt{1+\frac{64}{\hat{s}^{2}}-\frac{20}{\hat{s}}} - \frac{8}{\hat{s}} \right), \\
 y_{f/i} &=& \frac{3x+A_{x}-\hat{s}xA_{x}\pm\sqrt{4x(1-\hat{s}x)A_{x}+(\hat{s}xA_{x}-3x - A_{x})^{2}}}{2(1-\hat{s}x)}, \\
 z_{f/i} &=& \left[xA_{xy}+yA_{xy}-\hat{s}xyA_{xy}\pm\sqrt{4xy(x+y}-\hat{s}xy)A_{xy}\right. \nonumber\\ &+& \left. (\hat{s}xyA_{xy}-xA_{xy}-yA_{xy})^{2}\right]/\left[2(x+y-\hat{s}xy)\right],
\end{eqnarray}
where $\hat{s} = \frac{s}{m_{Q}^{2}}$.
\end{widetext}

\end{document}